\documentclass{aa}
\usepackage{graphics}
\begin{document}
\thesaurus{08(08.12.3; 08.12.2; 08.16.3; 10.07.2; 10.07.3)}          

\title{VLT observations of the peculiar globular cluster NGC\,6712, II:
luminosity and mass functions\thanks{Based on data obtained as part of 
an ESO Service Mode programme}}

   \subtitle{}

\author{G. Andreuzzi \inst{1}\inst{2} \and G. De Marchi \inst{2} \inst{3}
\inst{4} \and F. R. Ferraro \inst{5} \inst{2} \and F. Paresce \inst{2} 
\and L. Pulone \inst{1} \and R. Buonanno \inst{1}}

\institute{Osservatorio Astronomico di Roma, via di Frascati
33, I--00040 Monteporzio Catone, Rome, Italy
\and
   European Southern Observatory, Karl-Schwarzschild-Strasse 2, D--85748
   Garching, Germany 
\and    	
   Space Telescope Science Institute, 3700 San Martin Drive, Baltimore, 
   MD 21218, USA
\and
   affiliated with the Astrophysics Division, Space Science Department,
   European Space Agency
\and
   Osservatorio Astronomico di Bologna, Via Ranzani 1, I--40127, 
   Bologna, Italy
}

\offprints{G. De Marchi (demarchi@stsci.edu)}

\date{Received 16.9.2000; accepted 13.3.2001}

\titlerunning{Luminosity and mass functions of NGC\,6712}
\authorrunning{G. Andreuzzi et al.}
\maketitle

\begin{abstract}

We have carried out extensive VLT--FORS1 observations covering a fair
fraction of the intermediate metallicity globular cluster NGC\,6712 in
the $V$ and $R$ bands. We derive accurate colour--magnitude diagrammes
(CMD) and luminosity functions (LFs) of the cluster main sequence (MS)
for four overlapping fields extending from the centre of the cluster
out to a radius of $\sim 10^{\prime}$, well beyond the nominal tidal
radius, and for a control field at $\sim 42^\prime$ distance. The LFs
extend from the cluster turn-off (TO) at $M_R \simeq 4$ to the point at
which the incompleteness drops below $50\,\%$ (corresponding to $R
\simeq 23$ or $M_R \simeq 7.5$) for most fields studied. Cluster stars
become indistinguishable from field stars at $r \simeq 5^\prime$. The
shape of the cluster's LF and its variation with distance from the
centre in these ranges are well described by a standard multi-mass
static model having the following parameters: core radius $r_{\rm c} =
1^\prime$, half-light radius $r_{\rm hl} = 1\farcm8$, tidal radius
$r_{\rm t} = 5\farcm2$, concentration ratio $c=0.7$, and a power-law
global mass function (MF) with index $\alpha \simeq 0.9$ for masses
smaller than $0.8$\,M$_{\odot}$, i.e. for all detected MS stars, and
$\alpha\simeq-2.35$ for evolved objects. The MF obtained in this way is
consistent with that found in a preliminary investigation of this
cluster with the VLT Test Camera and confirms that this is the only
globular cluster known so far for which the global MF drops with
decreasing mass below the TO. Possible reasons for this unique
characteristic are discussed with the most likely associated with its
extreme vulnerability to tidal disruption.


\end{abstract}

\section{Introduction}

NGC\,6712 ($\alpha = 18^h\; 53^m\; 04.3''$, $\delta = -08^o \;42'
\;21.5''$) is a small and sparse globular cluster of intermediate
metallicity (concentration ratio $c=0.9$ and [Fe/H]$=-1.01$; Harris
1997).  Cudworth (1988), in his astrometric and photometric study
reaching down to just above the main sequence TO, finds that it is a
halo object in spite of its moderately high metallicity. One
interesting characteristic of this cluster is the presence in the core
of the high luminosity X-ray source (X1850-086) with an optical
counterpart (Anderson et al. 1993). This is unexpected for such a
loose cluster because most clusters with such sources tend to have a
much higher central concentration. To explain this peculiarity,
Grindlay et al.  (1988) have suggested that the cluster may be
currently re-expanding following the phase of core collapse when
densities were high enough to allow these binaries to be formed. Our
discovery (Ferraro et al., 2000) of the presence in the core of another
close binary, a $UV$- and $H\alpha$-excess object, most likely a
quiescent LMXB or a CV, only adds to the mystery.

Another, possibly connected and potentially even more interesting facet
of this cluster's structure is the fact that the first observations of
its MS taken by the VLT during its commissioning period and reported by
De Marchi et al. (1999) show a remarkable property of its MF near the
half-light radius. This is a clear and continuous drop with decreasing
mass starting already at the TO and continuing down to the observation
limit at $\simeq 0.5$\,M$_{\odot}$. MFs determined from LFs obtained
near the half-light radius are expected to faithfully reflect the shape
of the cluster's global MF (De Marchi et al.  2000; Vesperini \& Heggie
1997). For all the other clusters surveyed so far with HST in this mass
range, the global MF increases steadily with decreasing mass (Paresce
\& De Marchi 2000).

As suggested by De Marchi et al. (1999), this may be due to the fact
that its Galactic orbit forces the cluster to penetrate deeply into the
bulge. With a perigalactic distance smaller than 300\,pc, this cluster
ventures so frequently and so deeply into the Galactic bulge (Dauphole
et al. 1996) that it is likely to have undergone severe tidal shocking
during the numerous encounters with both the disk and the bulge during
its life-time. The latest Galactic plane crossing could have happened as
recently as $4\,\,10^6$\,year ago (Cudworth 1988), which is much smaller
than its half-mass relaxation time of 1\,Gyr (Harris 1996). It is
precisely on this basis that Takahashi \& Portegies Zwart (2000) have
suggested that NGC\,6712 has lost 99\,\% of its mass during its
life-time. And if the effects of this strong interaction have propagated
throughout the whole cluster and reached its innermost regions, as such
a gigantic mass loss implies, the peculiarly high density of core
binaries can be understood and justified for what would otherwise
appear an inconsequential cluster.

In an attempt to clarify this important issue and to better understand
the observable effects of tidal interactions, and especially to learn
more about the mechanisms leading to the dissolution of globular
clusters in the Galaxy and about the possible variation of the cluster
IMF with time in general, we have used the great power of the VLT and
the FORS1 camera to investigate in more detail the present structure of
NGC\,6712. The specific objective was to obtain a more precise LF of
the MS below the TO at various distances from the centre, so as to
evaluate the possible effects of mass segregation on the derived MF.
Another important objective was to sample more of the cluster at or
near the tidal radius to see whether or not one could detect an excess
of low mass stars ejected from the interior and still lightly bound to
the cluster and to correct for the numerous field stars expected in the
cluster field. In this paper, we report on the results of the analysis
of our VLT data set on NGC\,6712 with an emphasis on the cluster MS.
The analysis of the evolved part of the cluster CMD is the subject of
other papers (Ferraro et al. 2000; Paltrinieri et al. 2001).

\section{Observations and data reduction}

\begin{figure}[ht] 
\resizebox{\hsize}{!}{\includegraphics{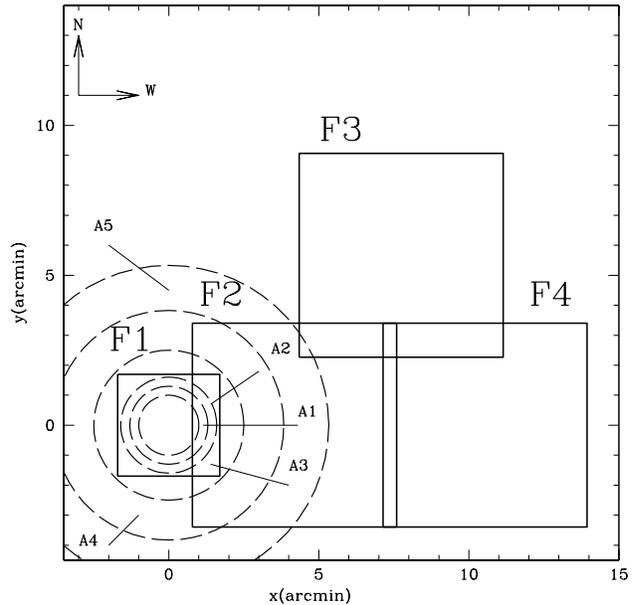}}
\caption{Locations of the four FORS1 fields on NGC\,6712. The centre
of the cluster is located at the origin of the coordinate system.
Dashed circles represent the annuli A1--A4 described in the text}
\label{Fig1}
\end{figure}

Our data consist of images of 5 fields in the $V$ and $R$ bands, four
of which are located as shown in Fig.\,\ref{Fig1}.  The fifth field,
used as a control field (field F0), is located $42^{\prime}$\,N of the
cluster centre and was imaged using FORS1 in its standard resolution
mode of $0\farcs2$/pixel.

\begin{figure}[ht] 
\resizebox{\hsize}{!}{\includegraphics{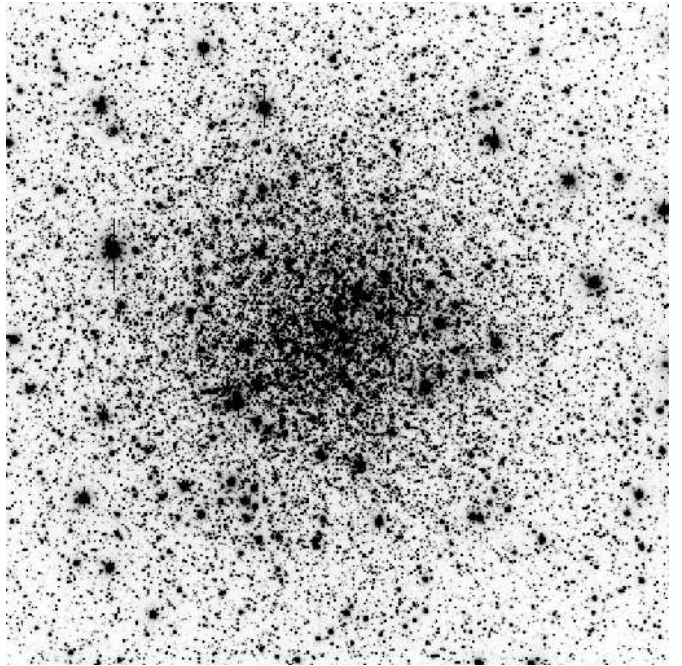}}
\caption{ VLT--FORS1 high resolution image (180\,s exposure) of
the core of NGC\,6712 in the $R$ band (field F1). The size of the image
is $3\farcm4 \times 3\farcm4$. North is up and East to the left} 
\label{Fig2}
\end{figure}

Because the level of crowding varies considerably from the core of the
cluster out to its periphery, our observations were carried out
according to the following strategy: the fields covering the external
regions of the cluster were imaged at low resolution (plate-scale
$0\farcs2$/pixel) and cover an area of $46.2$ arcmin square each
($6\farcm8 \times 6\farcm8$); they are located respectively
$5^{\prime}$\,W (field F2), $8^{\prime}$\,NW (field F3) and
$11^{\prime}$\,W (field F4) of the centre of the cluster; to improve
the photometry in the central regions, where the level of crowding is
particularly high, we have covered it with images taken in the high
resolution mode of FORS1 (plate-scale $0\farcs1$/pixel), with a field
$3\farcm4 \times 3\farcm4$ in size (field F1, shown in
Fig.\,\ref{Fig2}). To ensure a homogeneous calibration and to transform
the coordinates into a common local system from the centre of the
cluster out to the more external regions, each field has been selected
so as to overlap with at least a neighbouring one.

Since NGC\,6712 is situated in the midst of a rich star field at the
centre of the Scutum cloud (Sandage 1965), which is one of the highest
surface-brightness regions with high space-density gradients of
the Milky Way (Karaali 1978), we were anticipating that we would have
had to deal with significant foreground contamination and, therefore,
took the control field F0 in a region situated well away from the
cluster but representing a typical field in that area.

The journal of the observations is reported in Table\,1, where the
columns represent respectively: the fields covered, the date of each
observation, the distance of the fields from the centre of the cluster
and their coordinates, and the exposure time for the single images and
for each filter. Also listed in Table\,1 is the number of stars
detected simultaneously in both filters for each observed field (see
below).

\begin{table*}[ht]
\begin{small}
   \caption[]{Journal of the observations}
\[
\begin{array}{llrlllll}
                 \hline
                 \noalign{\smallskip}
{\rm Field} &   {\rm Date }& r & \alpha(2000) & \delta(2000)  & t_{R} (s)
                  &  \ t_{V} (s)  &  \ N_{obj}  \\
                 \noalign{\smallskip}
                 \hline
                 \noalign{\smallskip}
{\rm F1}    &   16.06.99  & 0   & 18^h53'04'' & -08^o 42'22'' & 
180 \times 4  & 180 \times 4 & 23057\\
{\rm F2}    & 15.06.99  & 5'  & 18^h52'45'' & -08^o 42'22'' & 
900 \times 2 & 900 \times 4     & 18777\\
{}    & 09.07.99  & {}  & {}             & {}         & 
900 \times 2    &{}     & {}\\
{\rm F3}    & 14.06.99  & 8'  & 18^h52'33'' & -08^o 36'42'  & 
900  \times 4    & 900  \times 4    & 16049 \\
{\rm F4}    & 08.06.99  & 11' & 18^h52'22'' & -08^o42'22''  & 
900  \times 4    & 900  \times 4    & 24092\\
{\rm F0}  & 20.06.99  & 42' & 18^h53'04'' & -08^o00'22''  & 
900  \times 4    & 900  \times 4    & 24117\\
                 \noalign{\smallskip} 
                 \hline
\end{array}
        \]
\end{small}
\end{table*}

Except for a small subset of the $R$-band images, we have adopted the
reduced and calibrated (i.e. bias-subtracted and flat-fielded) data as
provided by the standard ESO-VLT pipeline. Some of the raw $R$-band
data, however, had not been processed through the automated pipeline,
and for them we had to run standard IRAF routines following the same
recipe employed in the ESO-VLT pipeline. Subsequent data reduction and
analysis was done using standard IRAF photometry routines
({\it digiphot.daophot}).

Since our goal was to reliably detect the faintest object in these
images, for each field and filter we first created a mean frame using
all the applicable frames available and then ran the standard
digiphot.daofind routine on the average images so obtained to locate
the stars. Typical values of the PSF-FWHM are $0\farcs3$ and $0\farcs7$,
respectively at high and low resolution.  Although, in principle, we
could have also averaged images in different filters, the presence of
bad columns in the $R$-band frames (usually due to heavily saturated
pixels and spikes of the bright stars) suggested that we not follow this
approach. With a detection threshold set in the $V$ and $R$ band
typically at $3-5\,\sigma$ above the local average background level, we
obtained two independent coordinate lists for each field (one per
filter), which we then fed to the PSF-fitting routine {\em allstar} to
measure the instrumental magnitude of each object in each filter. We
found that a Moffat function gave the best representation of the shape
of the PSF, both at high and low resolution.

The positions of the identified objects in each mean $R$- and $V$-band
image were matched to one another, so as to obtain a final catalogue 
containing only the positions and magnitudes of the stars common to 
both filters.

Objects lying in overlapping regions between two adjacent fields were
used to determine the transformations between instrumental magnitudes
and to translate {\it local} frame coordinates to a {\it common}
coordinate system, with origin at the cluster centre. Typically, about
one hundred stars in each overlapping region were used to derive such
transformations. Only linear transformations were used to match
star measurements, with all magnitudes being referred to those of the
high resolution field (F1). For stars in the overlapping region,
multiple magnitude measurements were  averaged using appropriate
weights (which take into account the photometric quality of each
field). At the end of this procedure, a homogeneous set of
instrumental magnitudes, colours and positions (referred to field F1)
were obtained for a total of 106092 stars, in F1, F2, F3 and F4.

Instrumental (F1) magnitudes were finally transformed to the standard
Johnson system, using the stars in common with the bright portion of
the CMD which has been properly calibrated using 10 standard stars (see
Paltrinieri et al. 2001).

\begin{figure}[ht]
\resizebox{\hsize}{!}{\includegraphics{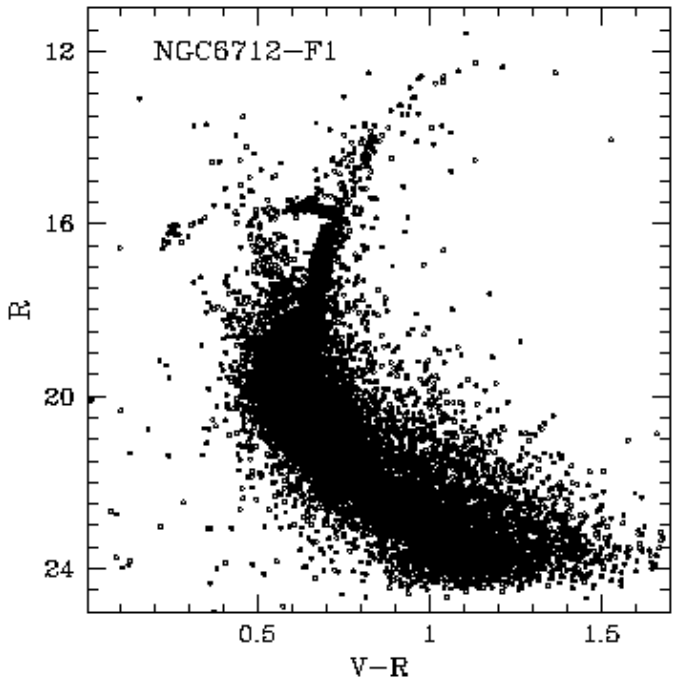}}
\caption{Color--magnitude diagramme of the stars in field F1 (core) of
NGC\,6712} 
\label{Fig3}
\end{figure}

Fig.\,\ref{Fig3} shows the total CMD of the central region of NGC\,6712
(field F1). The figure is obtained by merging the deep (180\,s long
exposure) and the bright (10\,s exposure) data covering the core of the
cluster.  In the following, we deal exclusively with the properties of
the cluster MS (from the TO at $R \simeq 19$ to $R \simeq 23$; see
Table\,1) while the bright portion of the CMD and the properties of the
evolved stellar population are discussed elsewhere (Ferraro et al.
2000; Paltrinieri et al. 2001).

\subsection{Incompleteness corrections}

A reliable assessment of the correction for incompleteness is a crucial
step because the main goal of this work is to compare with one another
the $R$-band LFs obtained at different radial distances from the centre
of the cluster and to extract in this way information on the underlying
global MF. Clearly, the completeness depends on the level of crowding
in the observed fields and, therefore, on their location with respect
to the cluster centre. In particular, an insufficient or inappropriate
correction for crowding will result in the distortion of the stellar
LF with a preferential loss of fainter stars and a relative increase of
bright and spurious objects. In our case, crowding is not the only
source of incompleteness: the distribution in luminosity of the stars
is also modified by the large number of hot pixels and bad columns
affecting the original images.

To correct our photometry for incompleteness, we ran artificial star
tests on both sets of frames ($V$ and $R$) independently, so as to be
able to estimate the overall completeness of our final CMD. First, we
applied the artificial star test to the mean $R$-band images:
artificial stars in each given $0.5$ magnitude bin were added
randomly to the frames, making sure not to exceed a few percent
($\leq 10\%$) of the total number of stars actually present in that bin
so as to avoid a significant enhancement of image crowding. We then
added an equal number of stars at the same positions in the $V$-band
frames and with a magnitude such that they would fall on the cluster
MS. It should be noted that we made the assumption that all artificial
stars were to lie on the MS since our intent was to verify the
photometric completeness of MS stars. This procedure was repeated for
all the bins of each field's CMD in both filters. To obtain a robust
result, we simulated more than 200,000 stars in 250 artificial images
for each field.
 
All pairs of $V$ and $R$ frames obtained in this way were then
subjected to the same analysis used for the original frames, with the
result being a catalogue of matching objects, each characterised by a
position and a pair of $V$ and $R$-band magnitudes. Each of these 250
catalogues (one per artificial pair of images) was compared with the
catalogue of input artificial stars: an artificial star was considered
detected only when its final position and magnitudes were found to
coincide with the input catalogue to within $\Delta x$, $\Delta y \leq
1.5$\,pixel, $\Delta$\,mag $\leq 0.3$. This approach allowed us to
build a map showing how photometric completeness varies with position 
in our frames.

If $N_{\rm rec}$ is the number of recovered stars in a given magnitude
bin, and N$_{\rm sim}$ the number of the simulated stars in the same
bin, the ratio $N_{\rm rec}/N_{\rm sim} = \Phi$ gives the completeness
in that bin for the location considered.

\subsection{Field subtraction}

In addition to correcting for photometric incompleteness, a reliable
determination of the LF of NGC\,6712 requires that we account for the
contamination caused by field stars. We have dealt with this correction
in a statistical way by using the comparison field F0, for which we
have produced a CMD and assessed photometric incompleteness precisely
as we did for all other fields. When it comes to measuring the LF --
our final goal -- we subtract from the stars found in a given magnitude
bin on the cluster CMD the objects detected in the same magnitude bin
in an area of equal size on the F0 field. Clearly, both numbers are
corrected for their respective photometric incompleteness before doing
the subtraction.

\begin{figure}[ht]
\resizebox{\hsize}{!}{\includegraphics{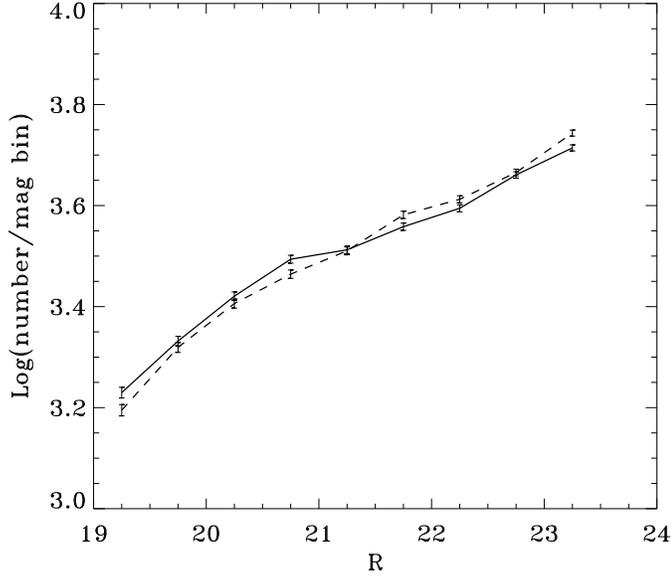}}
\caption{Luminosity functions measured in fields F0 (dashed line) and
F3 (solid line)}
\label{Fig4} 
\end{figure}

By applying the statistical field star subtraction described above, we
discovered that stars located in fields F3 and F4 can be considered as
belonging to the field because all the objects in the CMD of these
fields are statistically compatible with being field stars. We show
this in Fig.\,\ref{Fig4}, where we plot the $R$-band LF, corrected for
incompleteness, as measured in fields F3 and F0 (the solid and dashed
lines, respectively). The absence of any significant trend or
systematic departures of one function with respect to the other (to
within $2\sigma$) confirms that there are no residual cluster stars at
distances greater than $\sim 5^\prime$ from the cluster centre. As a
result of this finding, we decided to consider all stars lying in F3
and F4 as field stars, thus improving the statistical sample of the
field, and re-defined the decontamination procedure above using as a
comparison field the whole catalogue for F3, F4 and F0 ($r \geq
5^\prime$). This result also shows that the field around NGC\,6712 is
relatively uniform at our required level of accuracy and further
confirms that the statistical decontamination correction that we apply
is reliable.

We, therefore, regarded only the F1 and F2 fields as containing cluster
stars. Because of the richness of our sample of stars, we decided to
investigate the variation of the LF as a function of distance on a
scale smaller than the typical size of a frame. We have, thus, divided
our combined photometric catalogue into a number of annuli which are
centered on the cluster centre and whose size increases with distance
so as to guarantee a similar number of objects in each of them.
Table\,2 lists the five annuli A1 -- A5 that we have used, whose
positions are also marked in Fig.\,\ref{Fig1}. The columns in Table\,2
represent, respectively, the position and size of these annuli and the
number of objects measured before ($N$) and after ($N_S$) the
statistical decontamination. $N$ is the number of objects in the range
$19 < R < 22$ in each annulus before subtraction for field stars, while
$N_S$ is the number of objects after that subtraction in the same range
(selected so as to have photometric completeness always $> 50\,\%$ in
each annulus). Correction for photometric incompleteness is always
applied before performing the subtraction. We note here that annulus A3
is not properly annular in shape, in that it extends from a radius of
$96^{\prime\prime}$ all the way to the edge of field F\,1. The outer
radius given in Table\,2 ($105^{\prime\prime}$) is that which an
annulus of equivalent area would have.

\begin{table}[ht]
\caption{Location of the annuli A1 -- A5 in arcsec and in units of
r$_h$ ($1\farcm3$; Djorgovski 1993); see text for the definition of
$N$ and $N_S$}
\begin{tabular}{ccclr}
\hline
Annulus & \multicolumn{1}{c}{r/1"}  &  \multicolumn{1}{c}{$r/r_h$} &  
\multicolumn{1}{c}{$N$} & \multicolumn{1}{c}{$N_S$}  \\  
\hline
{\rm A1}    & 60 - 78   & 0.77 - 1.00 & 4335  & 3573\\
{\rm A2}    & 78 - 96   & 1.00 - 1.23 & 3742 & 2794\\
{\rm A3}    & 96 - 116  & 1.23 - 1.34 & 3433 & 2169\\
{\rm A4}    & 150 - 230 & 1.92 - 2.94 & 4434 &  744\\
{\rm A5}  &  230 - 320 & 2.94 - 4.1 &  4053 & 72 \\
\end{tabular}
\end{table}

We did not use the innermost regions of the cluster ($r \leq
60^{\prime\prime}$), as the high level of crowding and the large
ensuing incompleteness would have resulted in a poor determination of
the LF. Moreover, we did not include a region between annuli A3 and A4
because the level of crowding there is too high for the low resolution
of the FORS1 camera at $0\farcs2$/pixel and a standard seeing quality
of FWHM$\simeq 0\farcs6$.

\section{The luminosity function}

We have determined the $R$-band LF in each one of the annuli A1 -- A5
by counting the number of stars in bins of 0.5 magnitude each as a
function of the $R$ magnitude and correcting these values for
photometric incompleteness (see Sect.\,2 above). The same procedure
was applied to the comparison fields, and the number of stars found in
this way was subtracted from the LF of annuli A1 -- A5, after having
properly rescaled it so as to account for the different area covered
by the annuli and by the comparison field.

\begin{figure}[ht]   
\resizebox{\hsize}{!}{\includegraphics{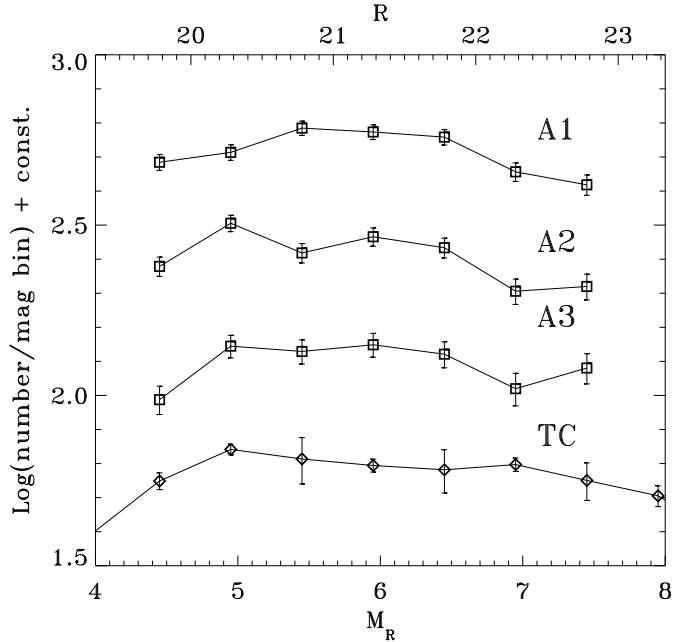}}
\caption{The LFs measured in annuli A1 -- A3, after correction for
photometric incompleteness and for field star contamination, are here
compared with the LF derived by De Marchi et al. (1999) with the VLT--TC
further out in the cluster (see text)}
\label{Fig5}
\end{figure}

Analysis of the LF of annuli A4 and A5 suggests that, in the region
covered by these annuli, the cluster population starts to become
negligible with respect to the field (see Table\,2), and the
completeness drops severely below 50\,\% at $R>22$. In particular, the
LF of annulus A5 oscillates statistically around zero. In the following
we, therefore, focus our investigation only on the region covered by
the first three annuli.  The LF obtained in this way in annuli A1 -- A3
are listed in Table\,3 and shown graphically in Fig.\,\ref{Fig5} (boxes).
Error bars in Fig.\,\ref{Fig5} reflect the total error associated with each
bin and include both the Poisson statistical error and the uncertainty
due to the correction for incompleteness:

\begin{equation}
\sigma_{bin} = \frac{\displaystyle \sqrt{N}}{\displaystyle \Phi} +
            \frac{\displaystyle N \sigma_{\Phi}}{\displaystyle \Phi^2}
\end{equation}
where $N$ is the number of stars before correction for incompleteness,
$\Phi$ and $\sigma_{\Phi}$ are the completeness factor and the
associated errors (ranging from $\sigma_{\Phi} \simeq 0.01$ to $\simeq
0.03$). We note here that the equation above provides a conservative
(upper limit) estimate of the uncertainty and is based on the formalism
developed by Bolte (1989). To account for the errors introduced by the
statistical subtraction of the comparison field, we have combined in
quadrature the $\sigma_{bin}$ of each annulus with that of the
comparison field.  It should also be noted here that, the very good
agreement between the field LFs shown in Fig.\,\ref{Fig4} and measured
$\sim 40^\prime$ away from one another suggests that our LFs are not
likely to be affected by local fluctuations in the distribution of
contaminating field stars.

\begin{table*}[t] 
\caption{
Luminosity functions for annuli A1, A2, A3, and for the stars belonging
to the control field in the $R$ band. In particular, mag-bin is the
centre of the bin used to obtain the LF; $N$ is the actual number of
stars observed; $\Phi$ is the completeness factor in each bin; $N_{\rm
C}$ is the number of stars after the correction for incompleteness;
$N_S$ is the number of stars after the subtraction for the control
field, properly rescaled; and $\sigma$ is the standard deviation of the 
associated uncertainty}
\begin{tabular}{cc|rrrrr|rrrrr|rrrrr} \hline
\multicolumn{2}{c}{mag-bin}&
\multicolumn{5}{c}{A1}&
\multicolumn{5}{c}{A2}&
\multicolumn{5}{c}{A3} \\ \hline%
$R$ & $M_R$ & 
$N$ & $\Phi$ & $N_C$ & $N_S$ & $\sigma$ & $N$ & $\Phi$ & $N_C$ & $N_S$ & $\sigma$ &
$N$ & $\Phi$ & $N_C$ & $N_S$ & $\sigma$ \\ \hline
%
19.75 & 4.42 &    530 &   0.91  &   582 &    484  &    25  &   462 &   0.92  &   502  &   379  &    24  &   359 &   0.88  &   407  &   244  &    23  \\
20.25 & 4.92 &    536 &   0.84  &   638 &    517  &    27  &   566 &   0.86  &   658  &   507  &    28  &   485 &   0.88  &   551  &   350  &    26  \\
20.75 & 5.42 &    585 &   0.78  &   750 &    609  &    29  &   496 &   0.84  &   590  &   415  &    27  &   486 &   0.85  &   571  &   338  &    27  \\
21.25 & 5.92 &    514 &   0.69  &   744 &    593  &    29  &   515 &   0.79  &   651  &   463  &    28  &   478 &   0.79  &   605  &   354  &    28  \\
21.75 & 6.42 &    433 &   0.58  &   746 &    573  &    29  &   465 &   0.72  &   645  &   430  &    28  &   477 &   0.77  &   619  &   332  &    29  \\
22.25 & 6.92 &    333 &   0.52  &   640 &    453  &    28  &   382 &   0.69  &   553  &   321  &    27  &   413 &   0.72  &   573  &   263  &    28  \\
22.75 & 7.42 &    315 &   0.50  &   630 &    415  &    28  &   359 &   0.60  &   598  &   331  &    28  &   415 &   0.63  &   658  &   302  &    30  \\
%
\end{tabular}
\end{table*}

Inspection of Fig.\,\ref{Fig5} immediately reveals that the three LFs,
measured at different radial distances from the centre of the cluster
in the range $60^{\prime\prime}$ to $105^{\prime\prime}$ ($0.77\,r_h$
to $ 1.34\,r_h$), are rather similar to one another: except for a
modest increase with decreasing luminosity up to R $\simeq 20.5$, they
are substantially flat or slightly decreasing down to $R \simeq 23$,
where the completeness drops below $\sim 50$\,\%.

An interesting check is to compare the LF in our annuli with that
derived by De Marchi et al. (1999) in a region of the cluster at a
slightly larger radial distance ($r = 135^{\prime\prime}$) and located
within our zone of avoidance (see Sect.\,2) between annuli A3 and A4.
This comparison is shown in Fig.\,\ref{Fig5}, in units of absolute
$R$-band magnitude. To convert our measurements from $R$ to $M_R$ we
have adopted a distance modulus in the $R$ band of $(m-M)_R =15.33$
(see Paltrinieri et al. 2001), which differs only marginally from the
value of $(m-M)_R = 15.31$ used by De Marchi et al. (1999). An
inspection of this figure confirms that for NGC\,6712 the LFs obtained
at different radial distances from the centre have a similar shape and,
over the magnitude range common to all of them, they are consistent
with one another within the quoted errors in the region from
$60^{\prime\prime}$ to $135^{\prime\prime}$.

This result is reassuring in that it confirms the validity of the
inverted MF found by De Marchi et al. (1999; see Sect.\,5 below). One
should note here that De Marchi et al. could not use a comparison field
to correct their LF for field star contamination and, because of this,
they suggest that the number of stars in their LF was likely to be
over-estimated at the faint end, since the field LF is expected to
increase steadily with magnitude. As it turns out, however, field stars
account for a small fraction of the total stellar population in these
regions, with cluster stars being from $\sim 4$ to $\sim 2$ times more
numerous at $R \simeq 21.5$ (respectively in annuli A1 and A3). This
fact, coupled with the slow rate of increase of the field LF with
magnitude (see Fig.\,\ref{Fig4}), makes field star contamination in the
TC field not sufficient to significantly alter the shape of the LF.

\section{The global mass function}

\begin{figure}[h]   
\resizebox{\hsize}{!}{\includegraphics{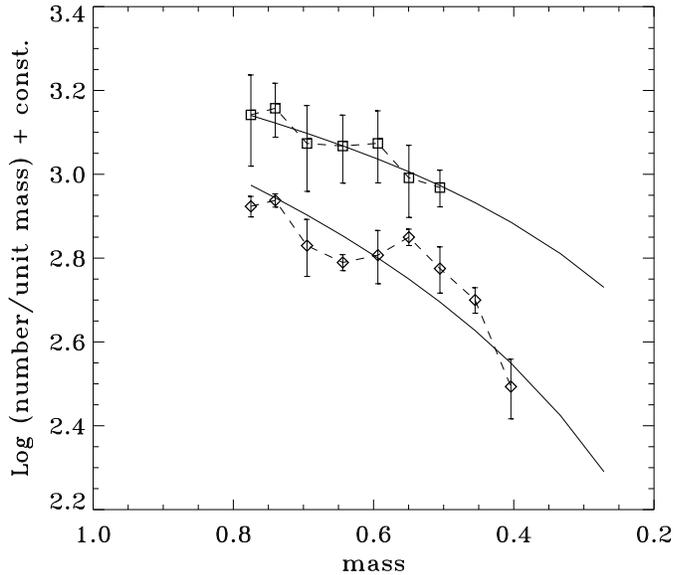}}
\caption{Comparison between the average MF in annuli A1 -- A3 (boxes)
and that derived by (De Marchi et al. 1999) with the VLT--TC (diamonds)
farther out in the cluster (see text). Thick solid lines mark the best
fitting power-law, with index respectively $\alpha = 0.9$ and
$\alpha=1.5$}
\label{Fig6}
\end{figure}

In Fig.\,\ref{Fig6}, we compare the local MF obtained in annuli A1 --
A3 using FORS1 with that measured by De Marchi et al. (1999) with the
TC at $r = 135^{\prime\prime}$. Rather than converting the three
individual LFs of annuli A1--A3 into MFs, we have first combined them
into one single function by averaging their values in each magnitude
bin, and have taken the standard deviation as a measure of the
associated uncertainty (error bars). We have done that for
compatibility with the approach used by De Marchi et al. (1999) for the
LF obtained with the TC, and because the LFs of the three annuli are
very similar to one another. In all cases (FORS1 and TC), we have
obtained the MF by dividing the corresponding average LF by the
derivative of the mass--luminosity (ML) relation appropriate for the
metallicity of NGC\,6712 ([Fe/H]$=-1.01$; Harris 1997). 
We plot here the results obtained using the ML relation of Baraffe et
al. (1997), but using the models of  Cassisi et al.  (2000) would have
yielded an almost identical result. The MFs derived in this way are
shown in Fig.\,\ref{Fig6}, where boxes connected by dashed lines
represent the average of the MF in annuli A1 -- A3 (upper curve) and in
the TC field (lower curve), whereas the thick solid lines show the best
fitting power-law MF, with index $\alpha=0.9$ and $\alpha=1.5$,
respectively for the FORS1 data (upper curve) and TC data (lower
curve). We should note here, however, that ignoring the data-point at
$\sim 0.4$\,M$_{\odot}$ in the lower curve would result in two MFs that
agree perfectly well with one another, within the errors, and the best
fitting power-law to the TC data would also have $\alpha \simeq 0.9$
(this agreement is hardly surprising as it was already implied by the
remarkable similarity of the LF). Regardless of the last data-point,
however, the net result in both cases is that {\em the number of stars
decreases steadily with mass}.

This result gives strong support to the claim of De Marchi et al. that
there is a relative deficiency of low mass stars with respect to the
stars at the TO ($M \simeq 0.75$\,M$_{\odot}$), although our data do not
reach deep enough to see whether this observed drop continues all the
way to $\sim 0.3$\,M$_{\odot}$ where all known GC feature a peak in
their MF (Paresce \& De Marchi 2000) before plunging to the H-burning
limit.

The dynamical evolution of a cluster depends on both the interaction
among the stars in the cluster, which locally modifies the distribution
of masses (internal dynamics, i.e. mass segregation), and on the
interaction with the Galaxy. In our particular case, even though
NGC\,6712 is likely to have experienced strong tidal shocks during its
life-time (De Marchi et al. 1999; Takahashi \& Portegies Zwart 2000),
one wonders whether the almost identical LFs (and ensuing MFs) that we
observe at the various radial distances as shown in Fig.\,\ref{Fig5}
can be ascribed to the internal two-body relaxation mechanism.

\begin{figure}[ht]   
\resizebox{\hsize}{!}{\includegraphics{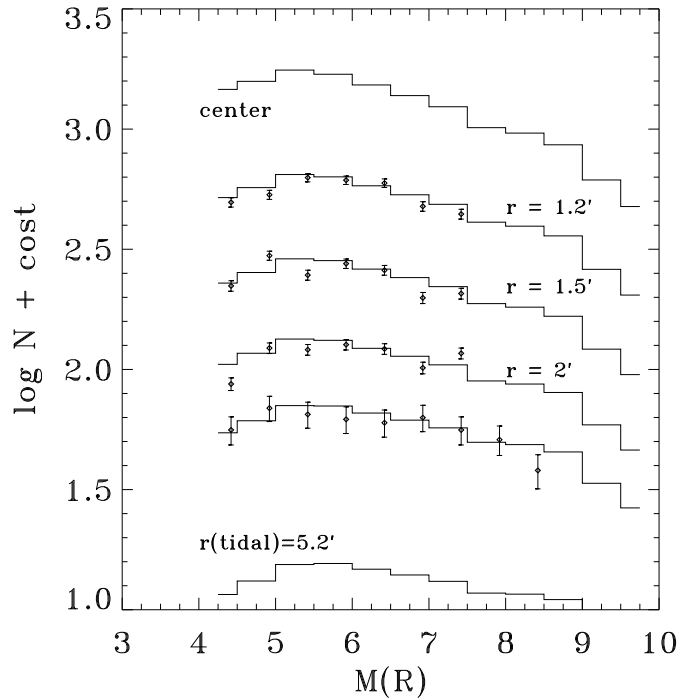}}
\caption{Theoretical LF as a function of distance as predicted by the
multi-mass Michie--King model described in the text. Diamonds represent
the observed LFs in annuli A1 -- A3 , and in the TC field}
\label{Fig7}
\end{figure}

To address this issue more specifically, we have simulated the
dynamical structure of the cluster using a multi-mass Michie--King
model constructed with an approach close to that of Gunn \& Griffin
(1979), as extensively described in Meylan (1987; 1988), and following
the technique developed more recently by Pulone et al. (1999) and De
Marchi et al. (2000), to whom we refer the reader for further details.

Each model is characterized by three structural parameters describing,
respectively, the scale radius ($r_c$), the scale velocity ($v_s$), the
central value of the gravitational potential ($W_o$), and a global MF
of the form $dN \propto m^{\alpha}$, where the exponent $\alpha$ would
equal $-2.35$ in the case of Salpeter's IMF. Stellar masses have been
distributed into nineteen different mass classes, covering MS stars,
white dwarfs (WDs) and other heavy remnants. All stars lighter than
$0.8 M_{\odot}$ have been considered still on their MS, while heavier
stars with initial masses in the range $8.5 - 100$\,M$_\odot$ have been
assigned a final mass of $1.4$\,M$_\odot$. WD have been subdivided into
three mass classes, following the prescriptions of Meylan (1987; 1988)
and assigned to the corresponding MS mass using the relations presented
by Weidemann (1988) and Bragaglia et al. (1995). The lower mass limit
is assumed to be $0.085$\,M$_\odot$. As already shown by Meylan (1987),
the exact value of different mass cutoffs does not significantly
influence the result of the dynamical modelling.

In order to fit the structural parameters of NGC\,6712, two different
mass function exponents have been adopted: $\alpha_{\rm up}$ for
stellar masses in the range  $0.8 - 100$\,M$_\odot$, and $\alpha_{\rm
ms}$ for MS stars below $0.8$\,M$_\odot$. We have furthermore assumed
complete isotropy in the velocity distribution.
 
From the parameter space, we have considered only those models
characterised by a surface brightness profile (SBP), a velocity
dispersion profile (VDP) and a mass-to-light ($M/L$) ratio which would
simultaneously agree well with their corresponding observed values. We
have further constrained the choice among the best fitting dynamical
models with the smallest reduced chi-squares, by imposing the condition
that the four observed LFs (A1--A3 and TC) had to be simultaneously
fitted by their theoretical counterparts as predicted by the mass
stratification of the dynamical structure of the cluster (see Pulone et
al. 1999 and De Marchi et al. 2000 for an extensive description of this
technique). As regards the SBP of NGC\,6712, in our simulations we have
followed the approach of Trager et al. (1985) and have used its
Chebyshev polynomial fit as evaluated on our own data (see
Fig.\,\ref{Fig8}), while for the VDP we have used the mean values
obtained by Grindlay et al.  (1988) within about three core radii. The
value of $M/L=0.7$ has also been taken from Grindlay et al. (1988).

Thanks to the many observational constraints that we force our model to
satisfy, we are able to considerably reduce the space in which
parameters can range. The best fitting set of parameters requires the
indexes of a power-law global MF $\alpha_{\rm up}$ and $\alpha_{\rm
ms}$ to take on values respectively around $-2.3$ (the Salpeter slope)
and $0.9$, as shown in Fig. 7. The line of sight velocity dispersion at
the centre is in this case $\sigma_{\rm v} = 4.3$\,km\,s$^{-1}$ and the
derived mass--luminosity ratio turns out to be $M/L=0.74$, in perfect
agreement with Grindlay et al.'s (1988) values $\sigma_{\rm v} =
4$\,km\,s$^{-1}$ and $M/L = 0.7$. The key result here, then, is that
the global MF of NGC\,6712 is indeed an inverted function, i.e. one
that decreases with decreasing mass below $\sim 0.8$\,M$_{\odot}$. The
other parameter values of the best fitting model are:  core radius
$r_{\rm c} = 1^\prime$, half-light radius $r_{\rm hl} = 1\farcm8$,
tidal radius $r_{\rm t} = 5\farcm2$, concentration ratio $c=0.7$, total
mass of the cluster $M_{\rm cl} = 7 \times 10^4$\,M$_\odot$, mass
fraction in heavy remnants $f=0.6$.
    
A large fraction of mass in the form of white dwarfs, neutron stars and
black holes such as the one that we find here might be surprising.
Heavy remnants are usually thought to account for up to 20 -- 30\,\% of
the mass of a cluster (Meylan \& Heggie 1997), whereas our result
suggests at least twice as many. The amount of heavy remnants depends
here exclusively on the shape of the IMF for stars more massive than
$0.8$\,M$\odot$. Because the latter have already evolved off the MS and
are no longer observable, the value of $\alpha_{\rm up}$ is not as
strongly constrained in our model as is that that of $\alpha_{\rm ms}$
for less massive stars. If, for instance, a value of $\alpha_{\rm up}=
-6$ were used, the fraction of heavy remnants could be brought down to
$\sim 25\,\%$.  Besides being highly suspicious in a mass range where
all known stellar populations display a Salpeter-like IMF (see e.g.
Kroupa 2001), such a steep MF exponent would also strongly affect the
$M/L$ ratio, forcing it to take on the value of $\sim 0.2$, which is
very discordant with the $M/L=0.7$ measured by Grindlay et al. (1988).
We, therefore, leave the value of $\alpha_{\rm up}$ unchanged and
consider in the next section the observational consequences that the
ensuing large fraction of heavy remnants implies.
 
The key result here, nevertheless, is that the current global MF of
NGC\,6712 is indeed an inverted function, i.e.  one that decreases with
decreasing mass, starting at least from $0.8$\,M$_\odot$. Although all
clusters whose LF has been studied in the core show an inverted local
MF there (as a result of mass segregation: see e.g. Paresce et al.
1995; King et al. 1995; De Marchi \& Paresce 1996), NGC\,6712 is {\em
the only known cluster so far} to feature an inverted MF on a global
scale. McClure et al. (1985) and Smith et al. (1986) have observed a LF
that drops with decreasing luminosity in the halo clusters E\,3 (see
also Veronesi et al. 1996) and Palomar\,5, respectively. The actual
shape of the corresponding global MF, however, is not known, as only a
single field is available in each cluster.  These objects are,
nonetheless, very interesting and should be studied using deeper,
higher resolution photometry at several locations in the clusters to
properly address the effects of mass segregation.

\section{Discussion and conclusions}

As Fig.\,\ref{Fig7} confirms, the observed (small) variation of the shape of the
LF with distance from the cluster centre is fully consistent with the
mechanism of mass segregation ensuing from energy equipartition as
currently understood in many other clusters (Meylan \& Heggie 1997).
We, therefore, cannot hold internal dynamical evolution responsible for
the observed inverted global MF in NGC\,6712: such a mechanism, in
fact, could only account for the inverted MF near the core of the
cluster, but not elsewhere (see De Marchi et al. 2000).

There remain, thus, only two ways to explain the inverted global MF of
this cluster, as already proposed by De Marchi et al. (1999). Namely,
either NGC\,6712 was born with an inverted IMF, at least for stars less
massive than $\sim 0.8$\,M$_{\odot}$, or the interaction with the tidal
field of the Galaxy (and more specifically disk and bulge shocking
through its frequent and repeated perigalacticon passages) has imparted
a strong modification to the stellar population of this cluster during
its life-time, as its orbit forces it to penetrate deeply into the
bulge at its disk crossings.

Although the former hypothesis cannot be completely ruled out, it is
highly unsatisfactory as it would explain one anomaly --- the inverted
global MF --- by invoking another one, namely an inverted IMF. A
careful investigation of the deep LF of a dozen GCs by Paresce \& De
Marchi (2000) shows no evidence of such an inverted IMF. Their sample
does, admittedly, only cover $\sim 10\,\%$ of the total population of
GC, but it contains clusters in widely different orbits and dynamical
states so that it can be regarded as representative of the whole
Galactic GC system. Still, the hypothesis of an inverted IMF cannot be
excluded, at least until the origin of the inverted LF observed long
ago in E\,3 (McClure et al. 1985) and Pal\,5 (Smith et al. 1986) is
understood (see Sect.\,4 above).

On the other hand, there seems to be more solid observational and
theoretical support for the latter hypothesis, namely that the cluster
has suffered severe tidal stripping which has remarkably altered its
stellar population. Recent calculations of the orbit of NGC\,6712 and
consequent destruction rate by Gnedin \& Ostriker (1997) and Dinescu et
al. (1999) clearly suggest that it has one of the highest destruction
rates of a large sample of Galactic GC and that it is one of the few
objects for which the tidal-shock rate is higher than its two-body
relaxation rate. One would, thus, expect that the strong tidal
interaction during the disk crossings and the consequent tidal shocking
should provoke a continuous loss of low-mass stars, especially from
beyond the half-light radius, and a consequent rapid change of the
stellar mass distribution. Precisely on this basis, Takahashi \&
Portegies Zwart (2000) have suggested that NGC\,6712 has lost 99\,\% of
its mass during its life-time and that it is now obviously only a pale
remnant of its initial much more massive condition.

\begin{figure}[ht]   
\resizebox{\hsize}{!}{\includegraphics{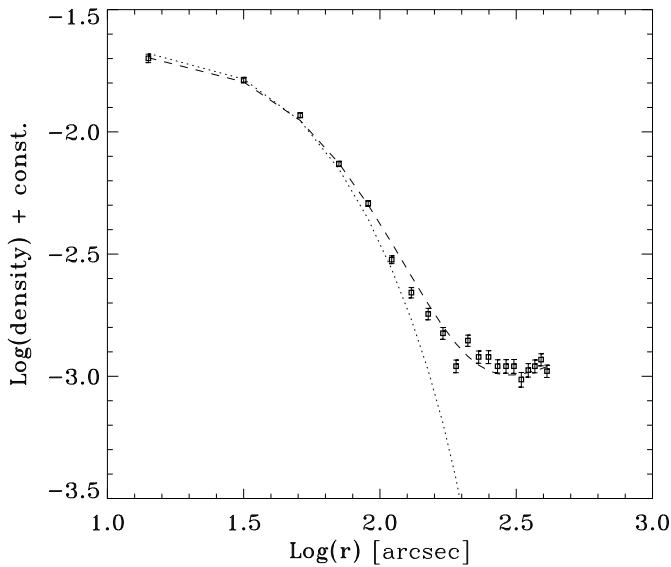}}
\caption{Surface density profile of $\sim 0.75$\,M$_{\odot}$ stars.
The thin line shows a King-type profile with $r_c = 1^\prime$
and $r_t=5\farcm2$, whereas the thick dashed line shows the
superposition of the latter on a plateau of field stars of uniform 
surface density} 
\label{Fig8}
\end{figure}

One might wonder, however, how the internal dynamics of a cluster which
has suffered such a tremendous mass loss could still conform so well to
the predictions of standard two-body relaxation and energy
equipartition as Fig.\,\ref{Fig7} indicates. Johnston et al.  (1999), on the
other hand, have shown that the tidal stripping operated by the Galaxy
on a GC results in a steady differential loss of stars (light stars
being dislodged more easily than massive ones) with the consequent
continuous decrease of the exponent $\alpha$ of the global MF and the
ensuing flattening of the latter. Although the inclination of the
cluster's orbit determines the rate at which the MF exponent decreases,
all orbits with perigalacticon within a few kpc of the Galactic centre
are exposed to this erosion. For clusters such as NGC\,6712, whose
orbit is mostly contained within the disk (Dauphole et al. 1996), the
heating due to disk and bulge shocking is diluted over a long time,
comparable with the dynamical relaxation time of the cluster
itself.\footnote{We note here that the half-mass relaxation time of
$t_{\rm hm} \simeq 1$\,Gyr quoted by Harris (1996) is most likely an
over-estimate for NGC\,6712. A value of 1\,Gyr is approximately the
average $t_{\rm hm}$ for all GCs, and NGC\,6712 is far less populous
and smaller than the average Galactic cluster.} It is, thus, not
unreasonable that two-body relaxation can proceed almost undisturbed,
and it does so on a continuously varying mass spectrum.

A natural consequence of tidal stripping is the formation of tidal
tails surrounding the cluster (Grillmair et al. 1995). While the latter
might be relatively easy to identify around clusters on highly inclined
orbits and currently well away from the Galactic plane, looking for
extra-tidal populations around NGC\,6712 is very difficult, because of
its orbit and current location in the Galaxy, and even more so because
the surface brightness of this excess of stars which might have been
ejected from the interior but are still loosely bound to it is expected
to be about 4 orders of magnitude lower than in the core (Johnston et
al. 1999). We have, nevertheless, searched the region near the
cluster's tidal radius ($ > 5^\prime$) but, not surprisingly, the
radial density profile that we have measured does not reveal any
obvious over-density near the cluster's boundary (see
Fig.\,\ref{Fig8}). On the other hand, we were forced to limit our
investigation to stars in the range $19.5 < R < 20.5$ (i.e. to $\sim
0.75$\,M$_{\odot}$ stars), so as to minimise the effects of variable
photometric completeness and crowding with distance, and it is quite
likely that most of these stars should today dwell preferentially in
the central regions of the cluster, rather than in its periphery, as a
result of mass segregation.

The radial density profile that we show in Fig.\,\ref{Fig8}, however, has
allowed us to define a more reliable tidal radius for this cluster. The
thick dashed line marks a typical King-type profile with $r_c =
1^{\prime}$ and $r_t=5\farcm2$, superimposed on a plateau of field
stars. A tidal radius of $\sim 5^\prime$ is fully consistent with our
finding of a statistically null cluster LF in annulus A5 (which extends
to $r=5\farcm1$). Although we have limited our analysis to stars for
which photometric completeness is always $> 85\,\%$, severe crowding
and the concentration of many saturated stars in the innermost regions
could make our determination of the core radius $r_c$ uncertain. Using
shorter FORS1 exposures of the central $\sim 2^\prime$ radius of this
cluster, however, Paltrinieri et al. (2001) also find a value of
$r_c\simeq 1^\prime$, in excellent agreement with that estimated here.

Thus, assessing whether NGC\,6712 was indeed much more massive in the
past than it is now, as suggested by the work of Takahashi \& Portegies
Zwart (2000), would require a more accurate search for tidal tails
surrounding the cluster, using a large field of view and sophisticated
reduction techniques such as those developed by Grillmair et al. (1995)
and, more recently, by Leon et al. (2000). On the other hand, the
severe field contamination would necessarily limit the effectiveness of
this technique. Moreover, even revealing the presence of tidal tails
would not provide strong constraints on the original cluster mass. To
be sure, tidal stripping has taken place throughout the whole life of
the cluster and the majority of the stars lost in this way should be
today totally unbound and dispersed elsewhere in the Galaxy.

A more precise, quantitative estimate of the original cluster mass
could, however, come from a census of the WD population in its core. If
NGC\,6712 was indeed originally as massive as $10^7$\,M$_\odot$, then a
large number of WDs should now populate its core. Even ignoring the
effects of mass segregation (which would further increase the WD
population in the core by draining them from the periphery and forcing
them to drift there by virtue of their higher mass), the prescriptions
of Renzini (1985) suggest that $\sim 2\,000$ WDs brighter than $M_V
\simeq 14$ ($V \simeq 28.5$) should dwell in the central
$100^{\prime\prime}$ radius of the cluster, if the latter had its
present mass $\sim 5$\,Gyr ago. Clearly, if its mass that long ago was
even only ten times as much as it is now, we would expect of order
$20\,000$ WDs within a $100^{\prime\prime}$ radius. Already on the basis
of the available data we can conclude that a large fraction ($\simeq
60\,\%$ or more) of the cluster's mass must be in the form of heavy
remnants (see Sect.\,4). Whether the cluster was originally only a few
times more massive than it is now or whether it was one of the most
massive in the Galaxy cannot be determined with certainty at present,
although the N-body simulations of Vesperini \& Heggie (1996) seem to
suggest the latter option. With a powerful instrument such as the
Advanced Camera for Surveys soon to be installed on board the HST,
however, this scenario can easily be tested observationally and the
suspected ongoing dissolution of NGC\,6712 reliably characterised.

\begin{acknowledgements}
We are indebted to Carlton Pryor (the referee) whose comments and
remarks have considerably strengthened the presentation of our results.
It is our pleasure to thank Isabelle Baraffe for providing us with the
tabulated theoretical M-L relations, and Barbara Paltrinieri for
carrying out the reduction of the short exposures of NGC\,6712. G.A.
gratefully acknowledges the hospitality of ESO through the Director
General's Discretionary Research Fund. F.R.F. gratefully acknowledges
the hospitality of the Visitor Programme during his stay at ESO when he
contributed to this paper. G.A., R.B., F.R.F. and L.P.  acknowledge the
finacial support of the Ministero della Universit\`a e Ricerca
Scientifica e Tecnologica through the programme "Stellar Dynamics and
Stellar evolution in Globular Clusters."

\end{acknowledgements}


\begin{thebibliography}{}
\bibitem{} Anderson, S., Margon, B., Deutsch, E., Downes, R. 1993, 
           AJ 106, 1049
\bibitem{} Baraffe, I., Chabrier, G., Allard, F., Hauschildt, P. 
           1997, A\&A, 327, 1054
\bibitem{} Bolte, M. 1989, ApJ, 341, 168
\bibitem{} Bragaglia, A., Renzini, A., Bergeron, P. 1995, ApJ, 443, 735
\bibitem{} Cassisi, S., Castellani, V., Ciarcelluti, P., Piotto, G., 
           Zoccali, M. 2000, MNRAS, 315, 679
\bibitem{} Cudworth, K. M., 1988A, AJ 96, 105
\bibitem{} Dauphole, B., Geffret, M., Colin, J., et al. 1996, A\&A, 313, 119
\bibitem{} De Marchi, G., Leinbundgut, B., Paresce, F., Pulone, L. 1999, AA 
           343, L9
\bibitem{} De Marchi, G., Paresce, F. 1996, ApJ, 467, 658
\bibitem{} De Marchi, G., Paresce, F., Pulone, L., 2000, ApJ 530, 342
\bibitem{} Dinescu, D., Girard, T., Van Altena, W. 1999, AJ, 117, 1792
\bibitem{} Djorgosvki, S., 1993, in ASP Conf. Ser. 50, eds. S. Djorgovski \&
           G. Meylan (San Francisco: ASP), 373
\bibitem{} Ferraro, F., Paltrinieri, B., Paresce, F., De Marchi, G. 2000, 
           ApJ, 542, L29
\bibitem{} Gnedin, O., Ostriker, J. 1997, ApJ, 474, 233
\bibitem{} Grillmair, C., Freeman, K., Irwin, M., Quinn, P. 1995, AJ, 109, 
           2553
\bibitem{} Grindlay, J.,Bailyn, C., Mathieu, R.,Latham, D., 1988, IAUS
           126, 659
\bibitem{} Gunn, J. E., Griffin, R. F., 1979, AJ 84, 752
\bibitem{} Harris W., 1996, AJ, 112, 1487
\bibitem{} Johnston, K., Sigurdsson, S., Hernquist, L. 1999, MNRAS, 302, 771
\bibitem{} Karaali, S., 1979, AAS, 35, 241
\bibitem{} King, I.R., Sosin, C., Cool, A. 1995, ApJ,452, L33
\bibitem{} Kroupa, P. 2001, MNRAS, in press (astro-ph/0009005)
\bibitem{} Leon, S., Meylan, G., Combes, F. 2000, A\&A, 359, 907
\bibitem{} Meylan, G. 1987, A\&A 184, 144
\bibitem{} Meylan, G. 1988, A\&A 191, 215
\bibitem{} Meylan, G., Heggie, D. 1997, A\&AR, 8, 1
\bibitem{} Paltrinieri, B., Ferraro, F., Paresce, F., De Marchi, G. 2001, 
	   MNRAS, in press (June 2001 issue), astro-ph/0102331
\bibitem{} Paresce, F., De Marchi G. 2000, ApJ, 534, 870
\bibitem{} Paresce, F., De Marchi G., Jedrzejevski, R. 1995, ApJ, 442, L57
\bibitem{} Pulone, L., De Marchi, G. Paresce, F., 1999, A\&A, 342, 440
\bibitem{} Renzini, A. 1985, Exp. Astr., 1, 127
\bibitem{} Sandage, A., Smith, L. L. 1966, ApJ 144, 886
\bibitem{} Takahashi, K., Portegies Zwart, S., 2000, ApJ, 535, 759
\bibitem{} Trager, S., King, I., Djorgovski, S. 1995, AJ 109, 218
\bibitem{} Veronesi, C., Zaggia, S., Piotto, G. et al. 1996, in ASP Conf. 
           Ser. 92, eds. H. Morrison \& A. Sarajedini (San Francisco: ASP), 301
\bibitem{} Vesperini, E., Heggie, D. 1997 MNRAS 289, 898
\bibitem{} Webbink, F. 1985, in Dynamics of star clusters, IAU Symp. 113, 
	   (Dordrecht: Reidel), 541
\bibitem{} Weidemann, V. 1987, A\&A, 188, 74
\end{thebibliography}
\end{document}